\documentclass[apj]{emulateapj}

\shorttitle{Gap Formation}
\shortauthors{Duffell \& MacFadyen}

\begin{document}

\title{Global Calculations of Density Waves and Gap Formation in Protoplanetary Disks using a Moving Mesh}

\author{Paul C. Duffell and Andrew I. MacFadyen}
\affil{Center for Cosmology and Particle Physics, New York University}
\email{pcd233@nyu.edu, macfadyen@nyu.edu}

\begin{abstract}

We calculate the global quasi-steady state of a thin disk perturbed by a low-mass protoplanet orbiting at a fixed radius using extremely high-resolution numerical integrations of Euler's equations in two dimensions.  The calculations are carried out using a moving computational domain, which greatly reduces advection errors and allows for much longer time-steps than a fixed grid.  We calculate the angular momentum flux and the torque density as a function of radius and compare them with analytical predictions.  We discuss the quasi-steady state after 100 orbits and the prospects for gap formation by low mass planets.

\end{abstract}


\section{Introduction}
\label{sec:intro}

It has been well over a decade since the discovery of ``Hot Jupiters", i.e. extrasolar planets orbiting their host stars at radii at which they are not expected to be able to form \citep{mq95}.  Yet the study of protoplanetary migration still poses fundamental theoretical challenges.  Migration is expected to be driven by gravitational torques exerted on the planet by the protoplanetary disk \citep{gt80}.  A major unsolved puzzle is that expected migration timescales are short compared to the lifetime of the disk (Type I Migration) \citep{gt79,gt80,w97,ar05,pt06,yl10}.  This suggests that a mechanism to halt planetary migration (or at least slow it down) is required, otherwise  planets are expected to migrate into their protostar or be kicked out of the system.  

Many mechanisms have been proposed to this end, including resonant interactions between multiple planets \citep{msg01}, disk turbulence, which might induce a stochastic component to migration \citep{np04}, radiative effects \citep{pm06}, and the effect of an inner disk cavity produced by the magnetic field of the central star \citep{shu94}.  In this work we focus on gap opening, which, when it occurs, is guaranteed to slow planet migration.  If a planet opens a gap in the disk, migration slows down to viscous timescales (Type II Migration) \citep{w97,np00,ol02,ss04,ar05,pt06,pn06,cm07}.  Depending on disk viscosity, this could significantly increase the time it takes for the planet to migrate.  It is therefore of interest to precisely determine what conditions are required for a planet to open a gap.

\begin{figure}
\epsscale{1.0}
\plotone{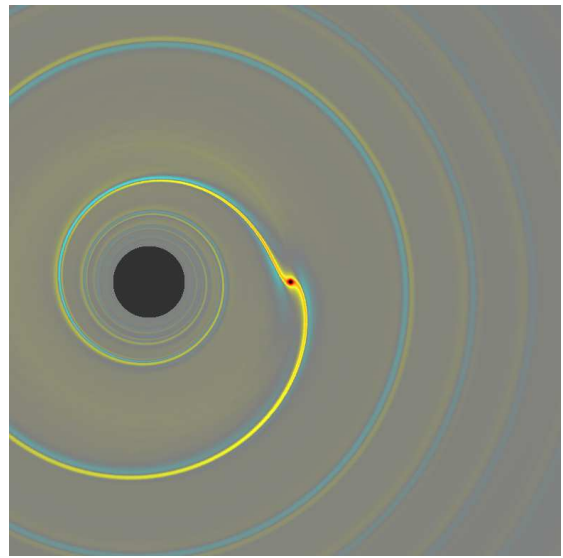}
\caption{ The perturbation to surface density caused by a low-mass planet, ${M_p = .0209 M_{Th}}$.  The planetary wake traces out a spiral shape.  This calculation used 4096 radial zones.
\label{fig:pretty} }
\end{figure}

The question of criteria for gap formation has been explored previously, e.g. \cite{wh89,lp93,gap,cm07}.  A gap is formed when gas is driven away from the planet, which requires an exchange of angular momentum between the planet and the gas in its neighborhood.  Merely exuding torque in the form of a spiral density wave (Figure \ref{fig:pretty}) is not sufficient -- this torque must actually be deposited in neighboring fluid elements, thus transporting angular momentum and driving disk evolution.  In the limit of a very low-mass perturber, the density wave driven by the planet will simply provide a uniform flux of angular momentum, without opening a gap.

On the other hand, if the perturbing mass is large enough to create a strongly nonlinear wave, the wave will quickly shock, dissipate, and transfer angular momentum between the planet and the gas, driving material away from the planet and forming a gap.  Between the two extremes there must exist a threshold mass for which a gap forms.  The threshold mass, ${M_{Th}}$, can be estimated by the condition for strong nonlinearity in the spiral wave \citep{lp93}:

\begin{equation}
M_p > M_{Th} = { c_p^3 \over \Omega_p G } = { M_* \over \mathcal{M}^3},
\label{eqn:thresh}
\end{equation}

where $c$ is the sound speed, $\Omega$ is the orbital frequency, $r$ is the orbital radius, and ${\mathcal{M} = r_p \Omega_p/c_p}$ is the Mach number, which for a thin disk is assumed to be equal to ${r/h}$, where $h$ is the disk scale height (the subscript "p" means the quantities are evaluated in the vicinity of the planet).  This threshold mass is also the mass for which the planet's Hill radius is of order the disk scale height.  If we assume the central star is a solar mass and the disk is orbiting at Mach 20, the threshold mass is ${41 M_\earth}$.  Alternatively, assuming a minimum mass solar nebula, the threshold mass is approximately ${21 M_\earth (r/1AU)^{3/4}}$ \citep{h81}.  In our work we use the former value (the two answers agree at ${r_p = 2.4 AU}$).  \cite{cm07} combined (\ref{eqn:thresh}) with a requirement that gap opening be faster than viscous filling of the gap:

\begin{equation}
1.1 ( {M_p \over M_{Th}} )^{-1/3} + 50 \alpha \mathcal{M} ( {M_p \over M_{Th}} )^{-1} \leq 1
\end{equation}
 
where $\alpha$ is the dimensionless viscosity, $\alpha = \nu/h c$.  These estimates give an upper bound; strong nonlinearity leads to gap opening.  However, the actual gap-opening threshold might be smaller than these estimates.  The basic idea behind (\ref{eqn:thresh}) is correct.  Gap opening requires not only the generation of a spiral wave, but a damping mechanism which transfers angular momentum from the wave to local fluid elements.  However, a point raised by \cite{gr01} is that there is damping even in the weakly nonlinear regime.  In other words, (\ref{eqn:thresh}) is a sufficient condition, but it might not be a necessary one.  A weakly nonlinear perturbation will also shock eventually, some distance from the planet.  Assuming negligible disk viscosity and that the wave shocks before reaching the edge of the disk, any massive perturber at a fixed radius is capable of eventually forming a gap, though for small enough perturbers the timescale is prohibitively long and Type I migration would occur too rapidly for gap opening.  Hence, for low mass planets, the formation of a gap depends on a competition between the timescale for gap opening and other relevant timescales, such as the viscous timescale or migration timescale.

In order to examine the details of gap opening by a weak shock, \cite{gr01} determined the wave form produced by a planet in the linear regime, then described its evolution into a shock assuming weak nonlinearity.  This semi-analytic result was calculated in the shearing box approximation, and then later took into account the cylindrical geometry of the system \citep{r02}.  For the latter case, some stronger predictions were made; in the global analysis, it was possible to take into account large-scale variations in density and sound speed in the disk, and to determine the conditions under which the density wave would leave the disk without shocking.

Recently, various groups have undertaken numerical investigations to confirm these results.  \cite{msi10} reported a dip in density (``partial gap") around planets of mass ${\sim .2 M_{Th}}$ using the shearing box approximation.  \cite{yl10} performed global simulations including planet migration and nonzero disk viscosity, demonstrating that ${\sim 10M_\earth}$ planets can have their migration halted if the disk viscosity is low enough.  However, \cite{drs1} pointed out that all of these results have focused on properties (like migration rates) which are derivative of the shock dissipation described by \cite{gr01}, rather than showing a converged calculation of the breaking wave itself.  They performed their own calculations at much higher resolution, focusing on smaller planetary masses and recovering the semi-analytical predictions to high accuracy.  They also noted \citep{drs2} that capuring the waveform properly requires much higher resolution than was attempted in previous work.  Their calculations used the shearing-box approximation; in the present work we present a similar result for the global case.

The global case is much more challenging numerically, for several reasons.  First, there is simply a much larger computational domain, which is not likely to be circumvented by using higher resolution near the planetary orbit, if the wave shocks far from the planet.  Secondly, the relevant dynamical timescales are sound-crossing timescales (${\sim r/c}$), but because protoplanetary disks orbit supersonically, the time-steps are generally Courant-limited by the orbital timescales at the innermost resolved orbit.  This can be orders of magnitude shorter, which means it may require a prohibitively large number of timesteps to reach a state resembling quasi-equilibrium.  Thirdly, and perhaps most importantly, because the disk is supersonic the bulk of the motion is pure advection across the grid, and underresolved advection errors can completely wash out subtle features of the motion (like a weakly nonlinear shock forming) taking place in the local Keplerian frame.  For these reasons, several hydrodynamic methods have been developed specifically for handling the challenges associated with global disk problems.  Examples include FARGO \citep{fargo} and RODEO \citep{rodeo}.

We perform our own calculations of proto-planetary disks using a new method, whereby instead of using a fixed numerical grid, we allow the computational cells to move and shear past one another with the bulk Keplerian flow.  The numerical method we use is a variant of the TESS code \citep{tess}, with several important modifications specifically designed for disk problems.  TESS uses moving finite volumes to solve the equations of gas dynamics in conservation form.  The motion of the cells is accomplished by performing a Voronoi tessellation of the computational domain each time-step.  However, the numerical scheme is completely specified for any kind of tessellation, so in principle the domain can be decomposed into whatever cell shapes are most advantageous.  In the present work, we choose to decompose the domain into wedge-like annular segments, as typically implemented for cylindrical ${(r,\phi)}$ grids (see Figure \ref{fig:disk}).  The cells  remain at fixed radii and rotate with the local angular velocity of the fluid.  As a result, the global calculations are effectively computed on a locally co-moving numerical mesh.

After reviewing relevant theoretical predictions from the literature in \S \ref{sec:anly}, and describing pertinent details of our numerical techniques in \S \ref{sec:num}, we present the results of our calculation in \S \ref{sec:res}, including a demonstration of convergence and a calculation of the global distribution of torque density and angular momentum flux, before summarizing in \S \ref{sec:sum}.

\section{Semi-Analytic Predictions}
\label{sec:anly}
 
Here we briefly summarize the theory regarding the generation of the spiral density wave and its subsequent nonlinear evolution.  For details, see \cite{gr01}, \cite{r02}.  We always work in the thin-disk approximation, where we ignore all vertically propagating modes, and the equations of motion reduce to Euler's equations in two dimensions:
\begin{equation}
\partial_t \Sigma + \partial_i ( v_i \Sigma ) = 0,
\end{equation}
\begin{equation}
\partial_t ( \Sigma v_j ) + \partial_i ( \Sigma v_i v_j + P \delta_{ij} ) = F_j,
\end{equation}
\begin{equation}
\partial_t ( E ) + \partial_i (  v_i ( E + P ) ) = F_i v_i.
\end{equation}
In the above, $\Sigma$ is the surface density, ${\vec v}$ is the fluid velocity, P is pressure, ${\vec F}$ is the total gravitational force including the fixed central mass of the protostar plus the orbiting planetary potential and $E$ is the fluid energy density,
\begin{equation}
E = {1 \over 2} \Sigma v^2 + \epsilon.
\end{equation} 
We use ${\epsilon}$ to denote the internal energy density of the fluid, which is related to the pressure via the equation of state,
\begin{equation}
P = (\gamma - 1) \epsilon,
\end{equation}
and $\gamma$ is the adiabatic index of the fluid.  In this work, we set ${\gamma = 1.001}$, effectively producing an isothermal equation of state.  We assume in this work that the central mass of the primary is fixed at the origin, and we do not take into account the corresponding force due to the accelerated coordinate system.

\subsection{Constant Density and Pressure}
The formulae for the semi-analytic predictions simplify if we assume a uniform background surface density and sound speed, and Keplerian velocity,
\begin{eqnarray}
\Sigma_0(r) , c_0(r) = \textrm{constant}, \\
\Omega_0(r) = \Omega_p (r_p/r)^{3/2}.
\end{eqnarray}
In this case, the linearized equations \citep{r02,ol02} predict that the spiral density wave generated by the planet will trace out a path described by:
\begin{equation}
\phi(r) = \phi_p - \textrm{sign}(r-r_p) \left( 3 - 2 \sqrt{r_p \over r} - {r \over r_p} \right) \mathcal{M}
\label{eqn:phi}
\end{equation} 
(this assumes a uniform sound speed, not a uniform mach number).  The perturbation can most conveniently be described using the following variables \citep{r02}:

\begin{equation}
\eta =  {3 \over 2 } \mathcal{M} \left( \phi + \textrm{sign}(r-r_p) \left( 3 - 2 \sqrt{r_p \over r} - {r \over r_p} \right) \mathcal{M} \right),
\label{eqn:eta}
\end{equation}

\begin{equation}
\chi = { \gamma + 1 \over 2 } { \delta \Sigma \over \Sigma_0 } \left( { \sqrt{2} (r/r_p) \over \mathcal{M} |  (r_p/r)^{3/2}  - 1 | } \right)^{1/2},
\label{eqn:chi}
\end{equation}

\begin{equation}
\tau = {3 \over 2^{5/4} } \mathcal{M}^{5/2} \left| {\int_1^{r/r_p} {  |s^{3/2}  - 1 |^{3/2} s^{-11/4} } ds} \right| .
\label{eqn:t}
\end{equation}

$\eta$ acts like an azimuthal coordinate, $\chi$ acts as a proxy for the density perturbation ${\delta \Sigma}$, and $\tau$ is essentially a radial coordinate, describing distance from the planet.  The integral in the formula for $\tau$ can be evaluated in terms of hypergeometric functions.  Close to the planet, in the shearing box approximation, these variables take on the following limiting behaviors:

\begin{equation}
\eta \rightarrow { 3 \over 2 }\left(  y/h + {3 \over 4} ({x/h })^2 sign(x) \right)
\label{eqn:eta0}
\end{equation}

\begin{equation}
\chi \rightarrow {\gamma + 1 \over 2 }\sqrt{2\sqrt{2} \over 3}{ \delta \Sigma \over \Sigma_0 } \left|{x \over h} \right|^{-1/2}
\label{eqn:chi0}
\end{equation}

\begin{equation}
\tau \rightarrow { 2^{3/4} \over 5 } (3/2)^{5/2} \left|{ x \over h } \right|^{5/2}
\label{eqn:t0}
\end{equation}

where ${y = r_p \phi}$ and ${x = r-r_p}$.  After re-scaling, these coordinates are identical to those used by \cite{drs1}.  Therefore, to make contact with these results, we define the rescaled parameters,

\begin{equation}
\eta' = {2 \over 3} \eta,~~ \chi' = {M_{Th} \over M_p} { \sqrt{ 3 \sqrt{2} } \over \gamma + 1} \chi,~~ \tau' = \tau {M_p \over M_{Th}}
\label{eqn:primes}
\end{equation}

When the density perturbation is expressed in terms of $\eta$, $\chi$, and $\tau$, the weakly nonlinear description of the system is governed by the inviscid Burgers' equation.  This is true both in the global case, and in the shearing box approximation.  Therefore, when described in terms of the coordinates $\eta'$, $\chi'$ and $\tau'$, the shock profiles computed globally should look identical to the shearing box profiles calculated in \cite{drs2}.

Using the known solutions to Burgers' equation, \cite{gr01} found an approximate formula for the location of the shock in terms of $\tau$,
\begin{equation}
\tau_{sh} = 1.89 + 0.79 {M_1 \over M_p} ,
\label{eqn:shock}
\end{equation}
where ${M_1 = (2/3)M_{Th}}$.

\subsection{Angular Momentum Flux}

Planets generate an effective viscosity in the disk, which can be characterized by the angular momentum flux
\begin{equation}
F_J(r) = \int_0^{2\pi} (\Sigma \delta v_{\phi} r) v_r r d\phi
\label{eqn:amf0}
\end{equation}
This can in turn be expressed in terms of a dimensionless function, $\Phi(\tau)$:
\begin{equation}
F_J(r) = {27 c_0^3 \Sigma_0 \over 2^{3/2} (\gamma+1)^2 \Omega } ({M_p \over M_1})^2 \Phi( {M_p \over M_1} \tau)
\end{equation}
\begin{equation}
\Phi(\tau) = \int \chi^2(\eta,\tau) d\eta
\end{equation}
Because ${\chi(\eta,\tau)}$ satisfies Burger's equation, ${\chi^2}$ is conserved until a shock forms.  After the shock has completely formed and its shape is described by an ``N-wave" \citep{ll59}, the integral falls off like ${\tau^{-1/2}}$.  Therefore, a reasonable approximate formula for the angular momentum flux is:
\begin{equation}
\Phi(t) = \left\{ \begin{array}
                  {l@{\quad:\quad}l}
                 \Phi(0) & \tau<\tau_{sh} \\
                 \Phi(0) \sqrt{\tau_{sh}/\tau} & \tau>\tau_{sh}
                  \end{array} \right.
   \label{eqn:Phi}
\end{equation}
In reality there is some smooth transition region, after the initial shocking time but before the shock has completely been converted into an N-wave.  The direct solution of Burger's equation by \cite{gr01} showed that the transition region adds a small correction to the appoximate scaling relation.

For radii at which the angular momentum flux (\ref{eqn:Phi}) is not uniform, angular momentum is being transferred from the density wave to the disk, which creates an effective viscosity.  Wherever the slope of (\ref{eqn:Phi}) is large, there will be disk evolution toward opening a gap.

\section{Numerical Method}
\label{sec:num}

\begin{figure}
\epsscale{1.0}
\plotone{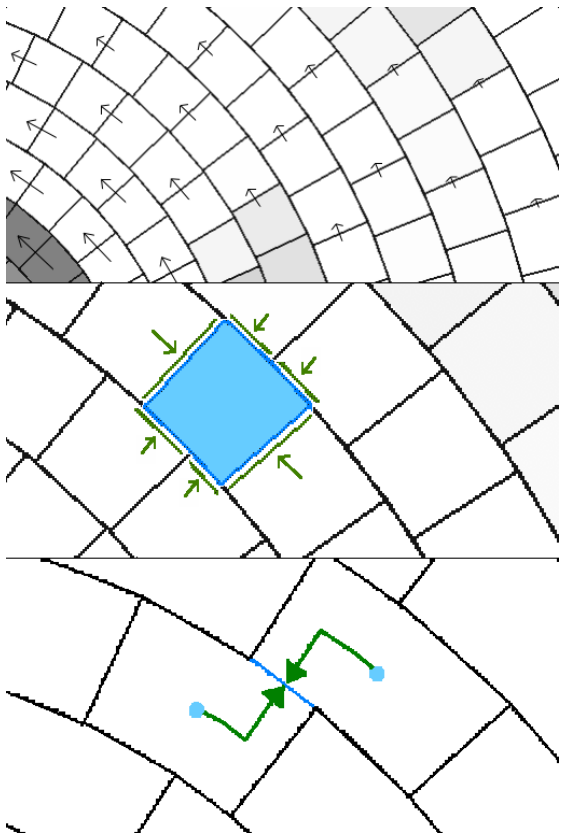}
\caption{ Top: The optional non-voronoi tessellation employed by Tess when solving problems involving gaseous disks.  Finite volumes are wedge-like annular segments which rotate independently at fixed radii.  Center: A close-up of one cell, indicating the six faces it shares with neighbors.  An approximate Riemann solver is used in the vicinity of each face to calculate the flux of each conserved quantity through the face.  Bottom: A schematic diagram depicting how primitive variables are extrapolated to each face, as input for ther Riemann solver.
\label{fig:disk} }
\end{figure}

\subsection{TESS code}
TESS is a finite volume hydrodynamics code constructed with a moving numerical mesh which can become effectively Lagrangian when the motion of its numerical cells is set equal to the local fluid velocity.  TESS is capable of solving general systems of hyperbolic equations in conservation-law form, in one, two, and three dimensions, with arbitrary coordinate geometry.  TESS is also a parallel code, configured to run on distributed-memory supercomputers, achieving parallelism via domain decomposition.  It is therefore capable of very high resolution calculations.  TESS is ideal for the study of fluid disks whose bulk motion is supersonic, especially for problems which need high accuracy which could potentially be compromised by large advection errors.  For more details on the numerical method, see \cite{tess}.  All of the features most recently added to the code (e.g. 3D and parallelization), will be detailed in a future publication.

Ordinarily, TESS accomplishes its mesh motion by creating a Voronoi tessellation of the computational domain to determine the shape and size of its finite volumes.  However, the TESS scheme is modular so that tessellations particularly suited to a given flow can be implemented.  For calculations involving proto-planetary disks, we choose to divide the computational domain into annular segments (similar to the tessellation which is performed by standard codes using polar coordinates) and allow the annular segments to each independently move with the local fluid angular velocity (Figure \ref{fig:disk}).  The numerical method relies on standard conservative finite-volume techniques.  The formulation requires the conservation-law form,
\begin{equation}
\partial_t U + \vec \nabla \cdot \vec F = S
\end{equation}
The finite-volume (integral) form of these equations is:
\begin{equation}
U^{n+1}_i dV^{n+1}_i = U^n_i dV^n_i + \Delta t ( - \sum_j \vec F_{ij} \cdot \vec {dA_{ij}} + dV^n_i S_i )
\end{equation}
where the U is the conserved quantity, F is the time-averaged flux through a face, S is a source term, and the sum is over adjacent faces.  This is simply a statement that the change in conserved quantities in a computational zone is equal to the time integral of the flux through all of its faces (in this 2D context, the "volume" of a zone is its area, and the "area" of a face is its length).  In this case the finite volumes are annular segments, which can have an arbitrary number of faces (usually about six in 2D), as indicated in the center panel of figure \ref{fig:disk}.

  Each face's time-averaged flux is calculated using an approximate Riemann solver.  The fluid quantities are extrapolated from each cell center to the center of each face using slope-limited gradients (bottom panel of Fig. \ref{fig:disk}):
\begin{equation}
P_f = P_c + ( \phi_f - \phi_c ) \partial_{\phi}P + ( r_f - r_c ) \partial_r P,
\end{equation}
where $P$ denotes a general primitive variable, the subscript ``$f$" means the variable is evaluated at the center of the face, and the subscript ``$c$" means it is evaluated at the center of the cell.  The gradients $\partial_{\phi}P$ and $\partial_r P$ are slope-limited gradients, calculated from the values of $P$ at neighboring cells.

  At the beginning of a timestep we assume a uniform state on either side of the face, which is valid close to the face.  The Riemann solver then determinees the time-averaged flux given this two-state problem.  For the moving faces, we modify the flux by subtracting off an advective term.  Once all of the fluxes are calculated, the rest of the time update behaves essentially like a standard logically-cartesian finite-volume code.  For further details on the method, see \cite{tess}.

Since the vast majority of the fluid motion is Keplerian, this scheme is very close to a Lagrangian formulation; there will be some advective fluxes in the radial direction, necessarily, but in principle we can always arrange, for example, to have the faces move in such a way that there is exactly zero advective flux in the azimuthal direction.  In the present work, we simply give each cell the local Keplerian orbital velocity.  This choice removes the bulk background flow and allows for computation of the fluid flow to take place on a grid effectively co-moving with the fluid.  As such, subtle flow features are captured which would be artificially diffused and dissipated if a fixed mesh were used.

The reason for this specialization in the case of disks is that while TESS is excellent at reducing diffusion due to large bulk motions, it can suffer from a small amount of numerical noise in the presence of large shearing motion; when two cells shear past one another, the face shared between the cells rotates quickly, producing noise.  While this noise is a very small price to pay for the vast reduction of diffusive fluxes, it is best to avoid it if possible.  The disk-tessellation into annular segments we have chosen for this study is an excellent solution to this problem.  Also, because it is a more restrictive geometry, the ``tessellation algorithm" is extremely efficient and straightforward to implement, when compared to a Voronoi tessellation.  We note that this idea is similar in spirit to orbital advection schemes like FARGO \citep{fargo}, but that the numerical method is clearly distinct.  FARGO is designed to work with the more restrictive logically cartesian grids, whereas our zones have a variable number of neighbors.

All the calculations were done using the basic TESS algorithm, but there are a few differences because we were able to take advantage of the simpler geometry.  The piecewise linear reconstruction has been modified for the annular segments, so that it looks exactly like a standard piecewise-linear method in the $\phi$ direction.  The other important distinction is that we formulate the hydrodynamic equations in terms of the angular momentum, so that angular momentum is explicitly conserved, as is desirable for disk calculations.

\subsection{Initial Conditions} 

All of our calculations are specified completely by the planetary mass as a fraction of the gap threshold mass ${M_p / M_{Th}}$ and the disk Mach number $\mathcal{M}$, which in this work will always be set equal to 20.  This value is consistent with observations of proto-planetary disks.  The initial conditions are given by:

\begin{eqnarray}
\Omega_0(r) = \Omega_p(r_p/r)^{3/2}\\
\Sigma_0(r,\phi) = \Sigma_p e^{-\Phi_p(r,\phi)/c_p^2}\\
c_p = \Omega_p r_p / \mathcal{M}\\
P_0(r,\phi) = c_p^2 \Sigma_0(r,\phi)/\gamma
\label{eqn:initial}
\end{eqnarray}

For simplicity, we choose a uniform density and pressure profile, though arbitrary profiles can be implemented straightforwardly.  Also for simplicity, we leave the planet at a fixed radius, though migration can easily be taken into account.  The exact value of ${\Sigma_p}$ is arbitrary, because the disk is not a source of gravity in our calculations; a change in ${\Sigma_p}$ is equivalent to a change of units.  Note also that our initial conditions have included a planetary atmosphere (an overdensity corresponding to the exponential in the formula for the surface density; ${\Phi_p(r,\phi)}$ is the planetary potential, described in the next section).  This choice was very important for reducing spurious transients, which can pose numerical issues.  It was noted by \cite{drs2} that orbital advection schemes like FARGO \citep{fargo} can potentially arrive at misleading results because these codes attempt to ``cheat" the Courant condition for their time-steps
\begin{equation}
\Delta t < \Delta x / \lambda_{max}
\end{equation}
by setting $\lambda_{max}$ to the sound speed rather than the orbital velocity (this is one of the primary motivations of FARGO).  The problem cited by Dong et al. is that there is also a gravitational timescale associated with accretion onto the perturbing planet, so that if a code's time-step criterion fails to take into account this additional timescale, qualitatively incorrect dynamics can result (e.g. spurious gap opening).  We have found that we are not subject to this additional timestep criterion, if we include the planetary atmosphere as an initial condition.  This is because we are searching for the quasi-steady-state solution, which should not have any features which evolve on the gravitational timescale.  Because we choose initial conditions including the planet's atmosphere which are close to stationary, the system can be evolved using the much longer timesteps associated with the sound crossing time of a computational zone.  We expect that the FARGO algorithm would also be able to advance using these larger timesteps, if given these initial conditions.  Note that this improves computational efficiency by at least a factor of the Mach number, which in this work is 20.  In fact, because the zones at the inner boundary have a much larger orbital velocity, the speed-up is actually about twice this factor.

\subsection{Planetary Potential}

Because the planet's position is inside the computational domain, it is not possible to use the exact (divergent) potential 
\begin{equation}
\Phi_p(s) = -{G M_p \over s},
\end{equation}
where $s$ is the distance from the planet.  Instead, it is standard practice to use an approximate potential with a smoothing length, ${\epsilon_s}$.  For two-dimensional disks, this smoothing length has a physical interpretation, because the gravitational force is vertically averaged.  As such, a smoothing length of ${\epsilon \sim 0.5h}$ is typically used \citep{ttw02,m02,mkm12}.  In our work we do not choose such a smoothing length because we are trying to make contact with results which apply in the limit ${\epsilon \rightarrow 0}$.  The potential should quickly converge to 1/r at distances larger than ${\epsilon_s}$, and so we use a potential which converges at nth order:
\begin{equation}
\Phi_p^{(n)}(s) = -{G M_p \over ( s^n + \epsilon_s^n )^{1/n} },
\end{equation}
where ${s = |\vec r - \vec r_p|}$ is the distance from the fluid element to the planet.  Typically we follow \cite{drs1} and use the fourth order version of this potential, ${n=4}$ and choose ${\epsilon_s/r_p = .005}$ (1/10 of a scale height), which is sufficient to produce the correct density waveform, as shown in section \ref{sec:linear}.  The form of our potential is shallower but less smooth, and was chosen for simplicity, so that it was straightforward to vary n.  This is not the same form as the fourth order potential used by \cite{drs1} which is deeper but smoother, and was chosen for stability.

\subsection{Boundary Conditions}

Boundary conditions are particularly challenging for astrophysical disk calculations.  There necessarily exists an inner boundary, at which the gravitational force due to the parent star is strongest, and numerical integrations are most prone to inaccurate cancellation between centrifugal and gravitational forces.  In general, if the system is allowed to evolve for long enough, spurious errors will develop at this inner boundary.  If the boundary condition there is not well behaved, these errors will eventually propagate to fill the rest of the computational domain.  Ideally, we would like a boundary condition that allows waves to simply propagate out of the inner boundary without reflecting back.
 
Rather than trying to directly implement a complicated inner boundary condition, an effective way to prevent reflections is to exponentially damp perturbations within some given radius ${r<r_{min}}$, as has been done by \cite{rodeo} and \cite{dvb06}.  Thus, waves propagating toward the inner boundary are completely damped and therefore unable to reflect back.  ${r_{min}}$ behaves as the boundary of the computational domain, and we interpret this damping as an outflow boundary condition at this radius.  As we shall see, the relevant dynamics are not negatively impacted by the simplifications we make at radii ${r<r_{min}}$.  Due to the good behavior of this boundary treatment, we decided to use the same boundary condition for large radii, ${r>r_{max}}$.  Typically we use the values ${r_{min} = 0.4 r_p}$, ${r_{max} = 3r_p}$, whereas our total computational domain extends to ${ 0.25r_p < r < 4r_p}$.

\section{Results}
\label{sec:res}

\begin{figure}
\epsscale{1.0}
\plotone{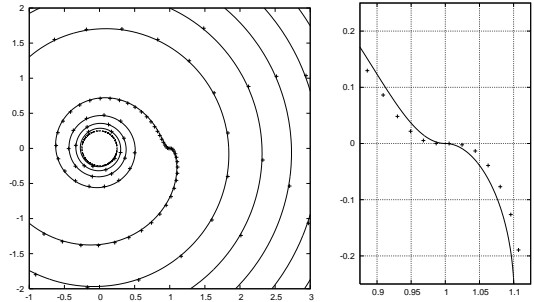}
\caption{ Position of the spiral density wave in the disk.  The curve plotted is the analytic prediction (\ref{eqn:phi}) which assumes a linear perturbation.  Points are density maxima in the numerical calculation.  The analytic formula is only valid far from the planet.  In the right panel, we show a close-up of the planet.  Here we see the same deviation from this curve as did \cite{drs1} (Paper I, Figure 2).
\label{fig:spiral} }
\end{figure}

\begin{figure}
\epsscale{1.0}
\plotone{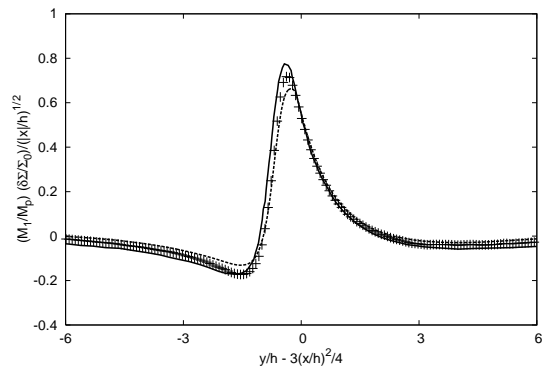}
\caption{ Density wave at ${\tau = 1.89 ~(r = .94 r_p) }$, for planet mass ${M_p = .0209M_{Th}}$ after 100 orbits, compared with the semi-analytical predictions of \cite{gr01}.  The solid curve is Goodman \& Rafikov's calculation, and the points correspond to our numerical measurement, using 14400 radial zones.  Variables x and y are shearing box coordinates, ${x = |r-r_p|,~y = r_p \phi}$.  The dashed line shows the results at much lower resolution, 2048 radial zones.  The linear phase is still reasonably well captured even at this much lower resolution.
\label{fig:linear} }
\end{figure}

\begin{figure*}
\epsscale{1.0}
\plotone{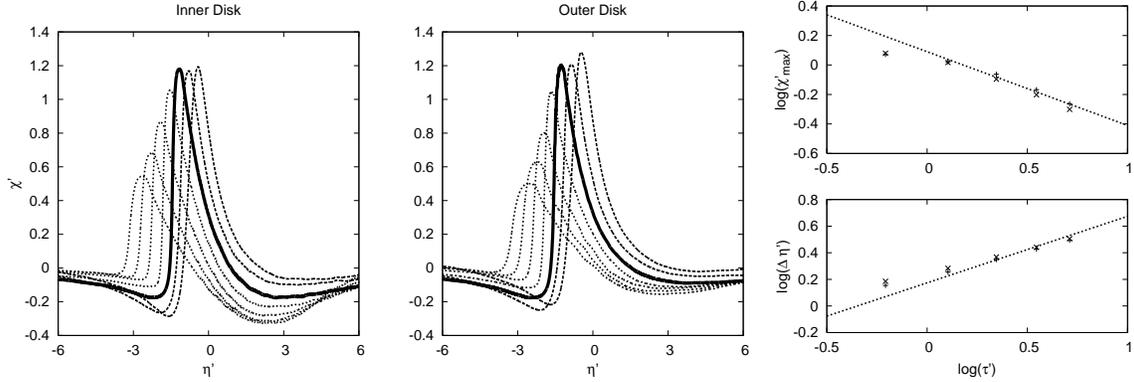}
\caption{ The nonlinear evolution of the density wave into a shock (${M_p = .0209 M_{Th}}$).  The left and center panels show the two separate shocks that form (inner disk and outer disk, respectively).  We plot the perturbation using the variables $\eta'$ and $\chi'$, equations (\ref{eqn:primes}).  The wave is shown at radii corresponding to ${\tau = 1.89, 10.8, 29.7, 60.8, 106,}$ and ${246}$, i.e. ${|r-r_p| = 1.27h (1.40h), 2.44h (2.94h), 3.51h (4.61h), 4.48h (6.45h), 5.39h (8.49h), 6.21h (10.72h),}$ and ${6.97h (13.2h)}$ in the inner (outer) disk.  The function ${\tau(r)}$ is given by equation (\ref{eqn:t}).  The bold curves describe the wave at ${\tau=29.7}$, where the shock is predicted to form.  The right panel shows the scaling relations for the width and the height of the shock.  The height ${\chi'_{max}}$ is given by the density maximum, and the width ${\Delta \eta'}$ is given by the magnitude of the ${\eta'}$ value at half-maximum.  This figure can be compared with Figure 1 of \cite{drs2}, where they performed the same calculation in the shearing box approximation.
\label{fig:shocky} }
\end{figure*}
 
The effect of resolution on the shock profile was extensively studied by \cite{drs1}.  They warn that this is a challenging problem which requires very high resolution and solid numerics in order to get it right.  The reason for this is that everything hinges upon calculating the profile of the linear density wave very accurately, and capturing the extremely weak shock which subsequently forms.  The shock profile is extremely sensitive to any form of viscosity, which includes numerical viscosity.  In fact, even to get the calculation right in the linear approximation using a discrete fourier transform, \cite{gr01} required 4096 x 8192 points in k-space.  Using the Athena code, \cite{drs1} were able to capture the dynamics accurately in the shearing box approximation, which required 3072 x 16384 grid cells.  We are performing a global calculation, which will require even higher resolution, if only because our computational domain is significantly larger.  This fact alone should increase the computational cost by roughly a factor of five in resolution.  Typically we use 14400 cells in the radial direction, and roughly ${2\pi}$ times this number in the azimuthal dimension.  Note that the number of azimuthal cells varies with radius, which is another feature of our method.  We choose our azimuthal resolution so as to keep a fixed cell aspect ratio of 1:1.  The total computational domain consists of roughly 500 million zones.  When varying the resolution, we found that the shock profiles were affected in the same way as was seen by \cite{drs2}, confirming their assertion that high resolution is important for this problem.

Because this is a challenging problem, we first demonstrate that our method accurately captures the predicted shock formation, essentially reproducing the shearing-box results of \cite{drs2}, but in the global case.

\subsection{Linear Phase and Shock Formation}
\label{sec:linear}

Equation (\ref{eqn:phi}) predicts the position of the spiral density wave, which we compare with our numerical calculation in Figure \ref{fig:spiral}.  To compare the analytic formula with our results, we measure the value of ${\phi}$ at which the surface density peaks for a given radius.  This is a relatively easy test, for which we found good agreement even at low resolution (for example, Figure \ref{fig:pretty} was computed at lower resolution, using only 4096 radial zones).  Very close to the planet, our result deviates from this prediction, in the same respect that was found by \cite{drs1}.  Actually, the density maximum is not predicted to coincide with this curve exactly; this would correspond to the density wave peaking at ${\eta=0}$, which is not predicted to occur by any theory.  However, we don't have to move far from the planet before the value of ${\eta}$ at which the density peaks is much smaller than the angular position ${| \eta_{max} | << \mathcal{M} \phi}$, meaning the position of the density maximum is generally a good proxy for ``wave position".  Regardless, the deviation close to the planet is expected and was also seen in the shearing box calculations.

Before nonlinear evolution sets in, the shape of the wave should agree with the linear calculations by \cite{gr01}, because the shearing box approximation is still appropriate in this regime.  Our calculation of the density wave agrees with linear theory (Figure \ref{fig:linear}) up to an overall normalization.

The most important test of our calculation is the accurate capturing of the shock, since this is the mechanism for exchange of angular momentum between the planet and the disk.  In figure \ref{fig:shocky} we track the shock formation.  Our results are consistent with the results of \cite{drs2} for this calculation.  We also confirm the scaling relations predicted by Burgers' equation for the amplitude and width of the dissipating shock as a function of $\tau(r)$ (Figure \ref{fig:shocky}, right panel).

\subsection{Torque Density}

\begin{figure}
\epsscale{1.0}
\plotone{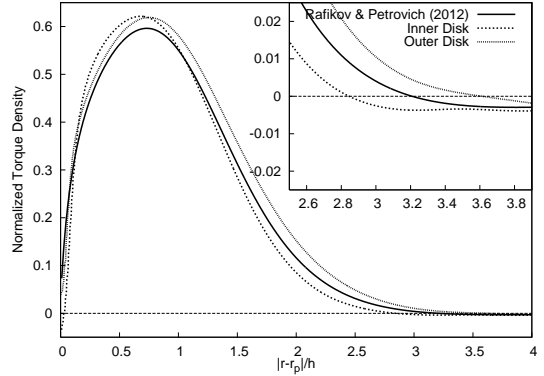}
\caption{ Normalized torque density for a planet of mass ${M_p = .0209 M_{Th}}$.  Normalized torque density is defined as ${dT/dr \times M_{Th} / ( \mathcal{M} \Sigma_0 G M_p^2 ). }$.  The solid curve is a linearized calculation by Rafikov in a shearing-box.  Dashed curves are our numerical result for the inner and outer disk.
\label{fig:torque} }
\end{figure}

\begin{figure*}
\epsscale{1.0}
\plotone{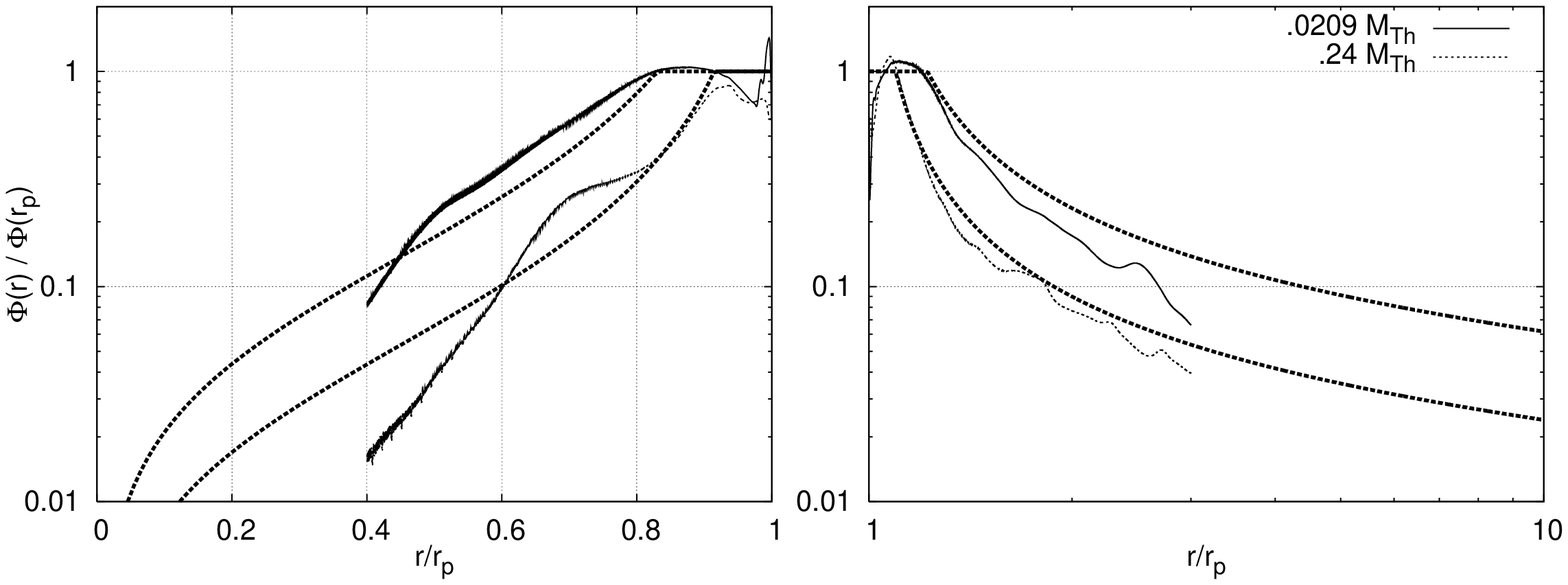}
\caption{ Normalized angular momentum flux ${\Phi(r)/\Phi(r_p)}$ for ${M_p = .0209 M_{Th}}$ and ${M_p = .24 M_{Th} ( 10 M_{\earth} )}$.  Thick dashed curves show the predicted scaling relation (\ref{eqn:Phi}).
\label{fig:amf} }
\end{figure*}

\begin{figure}
\epsscale{1.0}
\plotone{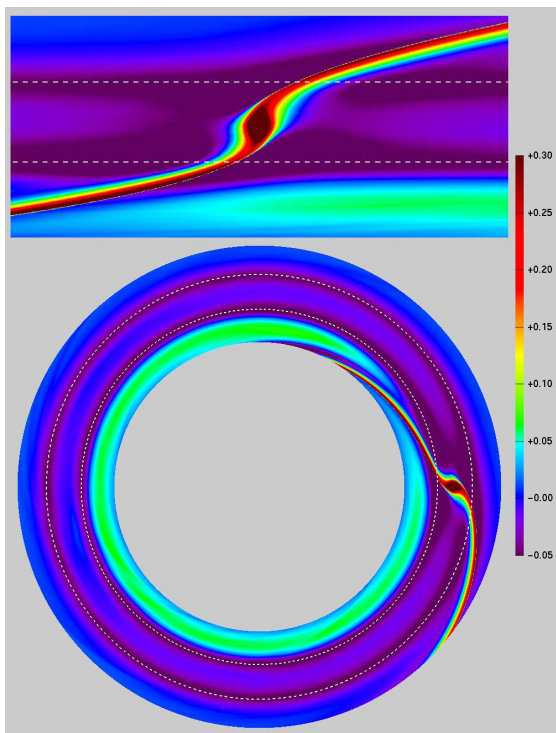}
\caption{ A ${10 M_\earth}$ planet giving hints of gap formation.  Color represents the perturbation to the surface density, ${\delta \Sigma / \Sigma_0}$.  Dotted lines are the theoretically predicted radii for shock formation, given by (\ref{eqn:shock}).  A downsampled subsection of the computational domain is shown, ${.75 r_p < r < 1.25 r_p}$.
\label{fig:gap} }
\end{figure}

\begin{figure}
\epsscale{1.0}
\plotone{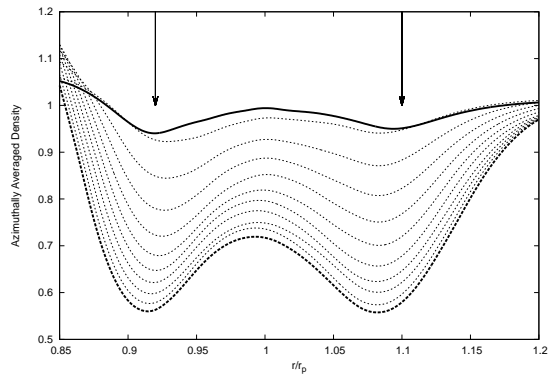}
\caption{ Azimuthally averaged density profiles for ${M_p = .24 M_{Th}}$, normalized to their initial value, at different times.  Arrows indicate the theoretically predicted shock positions.  The solid curve is our original calculation (same as Fig. \ref{fig:gap}) at 100 orbits.  All other curves were calculated at lower resolution (4096 radial zones) at 100 orbit intervals.  The thick dashed curve is the low resolution version at 1000 orbits.
\label{fig:lores} }
\end{figure}

Before reporting the results of the fully nonlinear calculation, we discuss the torque density in the linear perturbation.  Planetary migration timescales can  be calculated by adding up the total torque on the planet due to the net gravitational pull of the perturbed disk density profile.  In analytic calculations, this is typically done in Fourier space, but in this numerical study we can directly calculate torque density as a function of radius:
\begin{equation}
{dT \over dr} = \int \delta \Sigma r f_{\phi} r d\phi
\label{eqn:torque}
\end{equation}
where ${f_{\phi}}$ is the $\phi$ component of force per unit mass between the planet and the disk.

Early analytic results such as \cite{gt80} calculated the asymptotic behavior of the torque density, and found it to decay like a power-law.  Recently, however, this asymptotic behavior was shown by \cite{drs1} to be incorrect.  They found that the torque density changes sign at a particular distance from the planet,
\begin{equation}
r_{-} = r_p \pm 3.2 h
\end{equation}
which corresponds to
\begin{equation}
\tau_{-} = 17.
\end{equation}
\cite{r11} explained this disagreement by doing a more careful analytical calculation, without assuming that different Fourier harmonics were confined to the vicinity of Lindblad resonances.  Figure \ref{fig:torque} shows the torque density in our global calculations.  Our result is in agreement with that of \cite{r11}, though the amplitude is about a factor of three larger than 3D results \citep{dl10}.  As in other recent works, we observe a change in sign of torque density at finite distances from the planet, $|r-r_p| = 2.8h$ and $3.6h$, for the inner and outer disks, respectively.  Both of these radii are in agreement with the analytic value of $\tau_{-} = 17$ found in the shearing box case.  Actually, this particular result was not derived for the global case, but we find that expressing ${r_{-}}$ in terms of the $\tau$ coordinate gives a reasonable prediction for where the torque density becomes negative for both the inner and outer disk in the global case.

The difference between the torque in the inner and outer disk gives the net Lindblad torque on the planet.  Since we have a converged global calculation of torque density, we can numerically integrate it to find the net torque.  We find the net torque to be
\begin{equation}
T = - 2.0 ~ T_0,
\end{equation}
\begin{equation}
T_0 = r_p^4 \Omega_p^2 \Sigma_0 (M_p/M_*)^2 \mathcal{M}^2.
\end{equation}
It is worth noting that this result assumes uniform density and pressure in the disk, and would be modified significantly in the presence of nontrivial density and pressure profiles.  Many other torque calculations exist in the literature, some of which include these dependencies \citep{pp08,p10,ttw02,m11}.  This result is in qualitative agreement with others, though the coefficient does not quite agree with the semi-analytic 2D calculations by \cite{ttw02} and \cite{kp93}, who found ${T = 3.18 T_0}$ and ${T = 3.21 T_0}$, respectively.

\subsection{Angular Momentum Flux}

From our global numerical calculations we can confirm the prediction of global angular momentum flux as a function of radius \citep{r02}.  In Figure \ref{fig:amf} we show the function (\ref{eqn:amf0}) as computed from our data.  For comparison, we show the theoretical scaling relation (\ref{eqn:Phi}) calculated by \cite{r02}.  We appear to have better agreement with the semi-analytic theory in the outer disk than the inner disk, but the overall picture is clear; the angular momentum flux is roughly uniform between the planet radius and the radius at which the shock forms, after which shock dissipation causes the perturbation to deposit its angular momentum.  It should be noted that since (\ref{eqn:Phi}) is a scaling relation, we could get the initial waveform completely wrong and still have the correct (normalized) angular momentum flux, as long as the shock is correctly captured.  Predictions involving the angular momentum flux are therefore robust.

\subsection{Gap Formation}

Because the flux of angular momentum is not uniform, we see time-dependent disk evolution, which can lead to gap formation.  In Figure \ref{fig:gap}, we show the perturbation due to an intermediate-mass planet, ${.24 M_{Th} = 10 M_{\earth} }$, after 100 orbits.  Two gaps appear to be forming, one at the inner shock position and one at the outer shock position.

The question arises as to how long the disk evolution should continue.  This question proved computationally intensive at this resolution because the answer requires evolving the system for a large number of orbits.  To shed light on the situation, we performed the same calculation at much lower resolution (4096 radial zones), for 1000 orbits.  This result is shown in Fig. \ref{fig:lores}.  After 1000 orbits at this lower resolution, the planet hollows out a 45\% dip in density.  This is in qualitative agreement with the azimuthally averaged profiles of \cite{ll09}, though they assume a surface density gradient and include planet migration.  This result is not saturated; to evolve the system for long enough to see saturation we would have to go to even lower resolution.  The time-scale for gap opening is measured to be roughly 1500 orbits.  Regardless of what happens at late times, we have certainly found that a ${10 M_\earth}$ planet is massive enough to drive material away from it on a 1000 year timescale.  With negligible viscosity it is theoretically predicted that this trend will persist until the gap is very deep \citep{gap}.  \cite{msi10} saw a similar pattern in the shearing-box; some of their calculations suggested that gap opening could saturate by some mechanism (e.g. Rossby wave instability) after a few hundred orbits, though it was suggested by \cite{drs1} that these calculations might have have been underresolved.

We should note that this gap formation is the result of an accumulation of angular momentum flux over many orbits.  For this reason, it would not be possible for \cite{drs2} to have observed gap formation in their calculations, because they used an inflow boundary condition in their shearing box.  If they had used a periodic boundary condition, and if they had attempted using a ${10 M_\earth}$ planet, we expect that they would have seen the same gap that we have found in the global case.

\section{Summary}
\label{sec:sum}

We have performed high accuracy global protoplanetary disk calculations which extend the results of \cite{drs1} and have confirmed the analytical predictions of Rafikov for the case of a global disk.  Specifically, we directly measured the linear waveform, shock formation, and angular momentum flux, and find agreement with \cite{r02}.  We have also calculated the torque density for a low mass perturber and found agreement with recent shearing-box results \citep{drs1,r11} that find torque density to become negative a few scale heights away from the planet.

We have shown directly that a perturbing mass ${M_p < M_{Th}}$ (specifically, ten earth masses) can have a non-trivial impact on its environment, potentially leading to gap formation.  All of this, of course, assumes that the thin-disk approximation is valid to use here, which may not be true close to the planet, but it may be a reasonable assumption where the shock forms.  How much this assumption affects the result is unknown; this should be checked by three dimensional simulations.  We have  avoided speaking specifically about what conditions are necessary to open a gap.  To directly explore gap-opening criteria, we need to include planet migration and disk viscosity in our calculations, which is the topic for future work.

We have also demonstrated the power of using a moving mesh for performing hydrodynamical calculations involving gaseous disks, especially when a cylindrical grid geometry is used.  This result is the first of many planetary migration calculations using this method.  Now that we have demonstrated that it is possible to capture all of the important and subtle details of the shock formation and we have presented a converged global result, we can begin to study planet migration in this intermediate-mass range.  In addition, it is clear that this method should prove itself useful for studying many other subjects, for example binary black hole systems, the magnetorotational instability, and accretion onto compact objects.

\acknowledgments
This research was supported in part by NASA through grants NNX10AF62G and NNX11AE05G issued through the Astrophysics Theory Program and by the NSF through grant AST-1009863.  Resources supporting this work were provided by the NASA High-End Computing (HEC) Program through the NASA Advanced Supercomputing (NAS) Division at Ames Research Center.  We are grateful to Roman Rafikov, Jim Stone, Michael Kesden, and Jonathan Zrake for helpful comments and discussions.  We also thank the anonymous referee for his/her very thorough review.

{}

\end{document}